\documentstyle[preprint,prl,aps,epsf]{revtex}
\begin{document}
\title{Fourth Order Algorithms for Solving the Multivariable\\
 Langevin Equation and the Kramers Equation.
}
\author{Harald A. Forbert and Siu A. Chin}
\address{Center for Theoretical Physics, Department of Physics, 
Texas A\&M University, 
College Station, TX 77843}
\date{\today}
\maketitle
\begin{abstract}

We develop a fourth order simulation algorithm for solving the stochastic 
Langevin equation. 
The method consists of identifying solvable operators in the Fokker-Planck
equation, factorizing the evolution operator for small time 
steps to fourth order 
and implementing the factorization process numerically. A key contribution of 
this work is to show how certain double commutators in the factorization
process can be simulated in practice. The method is general,
applicable to the multivariable case, and systematic, with 
known procedures for doing fourth order factorizations. 
The fourth order convergence 
of the resulting algorithm allowed very large 
time steps to be used. In simulating the Brownian dynamics of 
121 Yukawa particles in two dimensions, the converged result of a first order 
algorithm can be obtained by using time steps 50 times as large. 
To further demostrate the versatility of our method, we derive two new classes
of fourth order algorithms for solving the simpler Kramers equation without
requiring the derivative of the force. The convergence of many fourth order
algorithms for solving this equation are compared.    

\end{abstract}
\section {Introduction}

A stochastic differential equation of the form
\begin{equation}
\dot x_i=G_i({\bf x})+D_{ij}\xi_j(t),
\label{sde}
\end{equation}
or its equivalent Fokker-Planck equation
\begin{equation}
{\partial\over{\partial t}} P({\bf x},t)=LP({\bf x},t)
\equiv\Bigl[{1\over 2}D_{ij}\partial_{i}\partial_{j}-\partial_i G_i({\bf x})\Bigr]P({\bf x},t),
\label{fkp}
\end{equation}
is used to describe a variety of physical and chemical processes\cite{risken}. 
Even in the Langevin case, where the diffusion matrix $D_{ij}$ is 
position independent, it is difficult to derive numerical algorithms for 
solving it beyond second order\cite{helf,drum,ukawa,chinnp,chinatm}. A direct 
Taylor expansion\cite{helf} approach
is laborious, giving no insight into
the overall structure of the algorithm and requires an 
eight term expansion to achieve 4th order accuracy\cite{eli}. Heretofore, no fourth 
order Langevin algorithm has been derived and applied to systems of more than
one particle.

The Fokker-Planck equation (\ref{fkp}) can be formally integrated to give
\begin{equation}
P({\bf x},t)={\rm e}^{tL}P({\bf x},0)=\Bigl[{\rm e}^{\epsilon L}\Bigr]^NP({\bf x},0).
\label{evol}
\end{equation}
This equation can be solved by factorizing the short time Fokker-Planck 
evolution operator ${\rm e}^{\epsilon L}={\rm e}^{\epsilon (T+D)}$ into exactly solvable
parts. In this work, we will take $D_{ij}=\delta_{ij}$ and define operators 
\begin{equation}
T={1\over 2}\partial_i\partial_i \qquad {\rm and}
\qquad D=-\partial_i G_i({\bf x}),
\label{tandd}
\end{equation}
with implied summations. This idea of operator factorization is 
not new, and has been used to derive a number of second order Langevin 
algorithms\cite{chinnp,chinatm}. We will briefly review the basic idea in
Section II. However, it is only recently that one learns 
how to factorize operators of the form ${\rm e}^{\epsilon (T+D)}$ to fourth order 
with positive coefficients\cite{sufour,chinlang}. All such fourth order 
factorizations require the evaluation of the double commutator $[D,[T,D]]$, 
which is rather formidable at first sight. We will show in Section III, how this 
commutator can be implemented judiciously to yield a fourth order Langevin 
algorithm. To demonstrate the high order convergence of this algorithm, we use 
it to simulate the Brownian dynamics of 121 Yukawa particles in two dimensions, 
a system that has been studied extensively by Branka and Heyes\cite{branka} 
using second order algorithms. 

To further demonstrate the utility of the factorization method for solving 
stochastic equations, we derive systematically a number of fourth order 
algorithms for solving the Kramers equation in Section IV.  
Drozdov and Brey\cite{drozdov} have used a similar factorization
method to solve this equation in one dimension using grid points. 
Hershkovitz\cite{eli} has also derived a fourth order algorithm by 
Taylor expansion. In both cases, it is not obvious how their respective
approaches can be generalized to the multivariable case. We give a detail 
comparison of all algorithms using Monte Carlo simulation, 
which can be easily generalized to any dimension. Finally, we summarize 
our findings and present some conclusions in Section V.

\section {Operator Factorization}

When the operator ${\rm e}^{\epsilon T}$ acts on $P({\bf x},t)$, it evolves 
the latter forward in time according to the {\it diffusion} equation
\begin{equation}
{\partial\over{\partial t}} P({\bf x},t)
={1\over 2}\partial_{i}\partial_{i}P({\bf x},t).
\label{tterm}
\end{equation}
If $\{x_i\}$ is a set of points distributed according to $P({\bf x},t)$, then
the distribution $\epsilon$ time later can be exactly simulated by 
updating each point according to 
\begin{equation}
x_i^\prime=x_i+\sqrt{\epsilon}\,\xi_i,
\label{stterm}
\end{equation}
where $\{\xi_i\}$ is a set of Gaussian distributed random numbers with zero 
mean and unit variance. When the operator 
${\rm e}^{\epsilon D}$ acts on $P({\bf x},t)$, it 
evolves the latter forward in time according to the {\it continuity} equation
\begin{equation}
{\partial\over{\partial t}} P({\bf x},t)
=-\partial_i [G_i({\bf x})P({\bf x},t)],
\label{tdterm}
\end{equation}
where ${G_i}({\bf x})P({\bf x},t)={J_i}({\bf x})$ is the probability current 
density with velocity field ${G_i}({\bf x})$. The continuity equation
can also be exactly simulated 
by setting 
\begin{equation}
x_i^\prime=x_i(\epsilon),
\label{trterm}
\end{equation}
where $x_i(\epsilon)$ is the exact trajectory determined by
\begin{equation}
{{d{\bf x}}\over{dt}}={\bf G}({\bf x}),
\label{traject}
\end{equation}
with initial condition $x_i(0)=x_i$.

Thus, if ${\rm e}^{\epsilon (T+D)}$ can be factorized into products of 
operators ${\rm e}^{\epsilon T}$ and ${\rm e}^{\epsilon D}$,
then each such factorization will give rise to an algorithm for 
evolving the system forward for time $\epsilon$. For example, the second order
factorization,
\begin{equation}
{\rm e}^{{1\over 2}\epsilon T}{\rm e}^{\epsilon D}
{\rm e}^{{1\over 2}\epsilon T}
={\rm exp}[\epsilon (T+D)+O(\epsilon^3)\cdots],
\label{ta}
\end{equation}
leads to a second order Langevin algorithm\cite{chinnp}
\begin{eqnarray}
y_i=&&x_i+\xi_i\sqrt{\epsilon/2}\,,\nonumber\\
x^\prime_i=&&y_i(\epsilon)+\xi^\prime_i\sqrt{\epsilon/2},
\label{altas}
\end{eqnarray}
where $\xi_i$ and $\xi_i^\prime$ are independent sets of
zero mean, unit variance Gaussian random numbers. For a second
order algorithm, it is sufficient to solve for the trajectory $y_i(\epsilon)$
correctly to second order in $\epsilon$, {\it e.g.} via a second
order Runge-Kutta algorithm:
\begin{equation}
y_i(\epsilon)=y_i+\epsilon G_i\left ({\bf	y}
+{1\over 2}\epsilon {\bf G}(\bf y)\right ).
\label{trk}
\end{equation}
Alternatively, one has the factorization,
\begin{equation}
{\rm e}^{{1\over 2}\epsilon D}{\rm e}^{\epsilon T}
{\rm e}^{{1\over 2}\epsilon D}
={\rm exp}[\epsilon (T+D)+O(\epsilon^3)\cdots],
\label{tb}
\end{equation}
which yields the second order algorithm
\begin{eqnarray}
y_i=&&x_i(\epsilon/2)+\xi_i\sqrt{\epsilon}\,,\nonumber\\
x^\prime_i=&&y_i(\epsilon/2).
\label{altbs}
\end{eqnarray}
Again, it is sufficient to solve the trajectory equations
$x_i(\epsilon/2)$ and $y_i(\epsilon/2)$ correctly to second order via
the Runge-Kutta algorithm. Despite the appearance that this algorithm 
requires solving the trajectory equation (\ref{traject}) twice,
it can be shown\cite{chinatm} that by expanding the two trajectories to 
second order and recollecting terms, one arrives at the second order 
Runge-Kutta Langevin algorithm\cite{helf,drum,ukawa}. However, the canonical 
form of (\ref{altbs}), with two evaluations
of the trajectory, usually has a much smaller second order error 
coefficient.

     The method of operator factorization thus appears to provide a
systematical way of generating higher order algorithms. Unfortunately, 
Suzuki\cite{nogo} proved in 1991 that, beyond second order, for any 
two operators, $T$ and $D$, it is impossible to factorize the evolution
operator as 
\begin{equation}
\exp[\epsilon (T+D)]
=\prod_{i=1}^N\exp[a_i\epsilon T]\exp[b_i\epsilon D] 
\label{facab}
\end{equation}
for any finite $N$, without having some coefficients $a_i$ 
and $b_i$ being negative. In the present context, since 
${\rm e}^{a_i\epsilon T}$ is the diffusion kernel, a negative $a_i$ 
would imply that one must simulate the diffusion process 
backward in time, which is impossible. Thus
factorizations of the form (\ref{facab}) cannot be used to derive 
higher order Langevin algorithms.
 
\section {A Fourth Order Langevin Algorithm}
 
The essence of Suzuki's proof is to note that in order to obtain a
fourth order algorithm, one must eliminate third order error terms involving
double commutators $[T,[D,T]]$ and $[D,[T,D]]$. With purely positive coefficients 
$a_i$ and $b_i$, one can eliminate either one or the other,  but not both. 
Thus to obtain a fourth order factorization with all positive coefficients,
one must retain one of the two double commutators. 
Recently, Chin\cite{chinlang} has derived three 
such factorization schemes, two of which were also found previously by 
Suzuki\cite{sufour}.

The form of the operators $T$ and $D$, as given in (\ref{tandd}), dictates 
that one should keep only the commutator $[D,[T,D]]$, which is at most a
second order differential operator. 
Since the velocity (or force) field ${\bf G}$ is usually given in terms of a
potential function $V({\bf x})$, 
\begin{equation}
G_i({\bf x})=-\partial_i V({\bf x}),
\label{vpot}
\end{equation}
the double commutator has the form
\begin{equation}
[D,[T,D]] = \partial_i \partial_j f_{i,j} + \partial_i v_{i},
\label{dcom}
\end{equation}
where
\begin{eqnarray}
f_{i,j}\,&&\equiv 
V_{i,j,k} V_{k} - 2 V_{i,k} V_{j,k} \nonumber \\
v_i\,&&\equiv {1\over 2} \left(
2 V_{i,j,k} V_{j,k} + V_{i,j} V_{j,k,k} - V_{i,j,k,k} V_{j} \right).
\label{fdef}
\end{eqnarray}
The indices on $V$ indicate corresponding partial derivatives.
Since the operator $D$ requires solving for the particle's trajectory, 
we must minimize its occurrence. This dictates that we use a variant of 
Chin's scheme B\cite{chinlang} to factorize 
\begin{eqnarray}
\exp\left[\epsilon \left( T + D\right) \right] &=&
\exp\left[ {\epsilon\over 2}\left(1-{1\over\sqrt{3}}\right) T \right]
\exp\left( {\epsilon\over 2} D \right)
\exp\left( {\epsilon\over\sqrt{3}}\tilde{T} \right)
\nonumber \\ &\times&
\exp\left( {\epsilon\over 2} D \right)
\exp\left[ {\epsilon\over 2} \left(1-{1\over\sqrt{3}}\right) T \right]
+O(\epsilon^5),
\label{dmc4p2}
\end{eqnarray}
where we have included the double commutator in $\tilde{T}$
\begin{equation}
\tilde{T}=T + {\epsilon^2\over 24}(2\sqrt{3} - 3)
[D,[T,D]].
\label{ttid}
\end{equation}
To obtain a fourth order algorithm, we must simulate this new
term 
\begin{equation}
\exp\left( {\epsilon\over\sqrt{3}}\tilde{T} \right)
=\exp\left[ {\epsilon\over\sqrt{3}} T 
+ {\epsilon^3\over 24}\left(2 - \sqrt{3}\right)
\left(\partial_i \partial_j f_{i,j} +
\partial_i v_i \right) \right]
\label{newt}
\end{equation}
correctly to 4th order. 
If we simply took all $x$ dependent terms in this operator as fixed, 
evaluated at the starting point, this operator would describe 
a non-uniform Gaussian random walk. However, this normal ordering 
would be correct only to third order. To implement 
it to fourth order, we first decompose it as
\begin{equation}
\exp\left( {\epsilon\over\sqrt{3}}\tilde{T} \right)
=
\exp\left( {\epsilon\over 2\sqrt{3}} T\right)
\exp\left[
 {\epsilon^3\over 24}\left(2 - \sqrt{3}\right)
\left(\partial_i \partial_j f_{i,j} +
\partial_i v_i \right) \right]
\exp\left( {\epsilon\over 2\sqrt{3}} T\right)+O(\epsilon^5).
\end{equation}
If $f_{i,j}$ is positive definite, normal ordering
the middle operator above, {\it i.e.} interpreting it as
a non-uniform Gaussian random walk with $f_{i,j}$ evaluated at the 
starting point, would be correct to 4th order (actually to 5th order). 
However, if some eigenvalues
of $f_{i,j}$ were negative, we would not be able to sample the
operator as a Gaussian walk. To avoid this possibility, 
we implement the normal order process as follows:   
\begin{eqnarray}
\exp&&\left( {\epsilon\over\sqrt{3}}\tilde{T} \right)
=
\exp\left( {\epsilon\over 2\sqrt{3}} T\right)
{\cal N} \left\{
\exp\left[
 {\epsilon^3\over 24}\left(2 - \sqrt{3}\right)
\left(\partial_i \partial_j f_{i,j} +
\partial_i v_i \right) \right]
\right\}
\exp\left( {\epsilon\over 2\sqrt{3}} T\right)
\nonumber\\
&&=																	   
{\cal N} \left\{
\exp\left[ {\epsilon\over 2\sqrt{3}} 
\left({1\over 2} \partial_i \partial_j \delta_{i,j}\right)
 +{\epsilon^3\over 24}\left(2 - \sqrt{3}\right)
\left(\partial_i \partial_j f_{i,j} +
\partial_i v_i \right) \right]
\right\}
\exp\left( {\epsilon\over 2\sqrt{3}} T\right),
\end{eqnarray}
where ${\cal N}$ denotes the normal ordering of all derivative operators
to the left. Since the left (and only the left) operator 
$\exp( {\epsilon\over 2\sqrt{3}} T)$ is already normal ordered with respect 
to the position-dependent operators in the middle term, the two normal 
ordered exponentials can be combined to remove the restriction of a positive 
definite $f_{i,j}$. Now, only the full covariance matrix $C$ needs to be 
positive definite, which will always be the case for $\epsilon$ sufficiently small.
The final normal ordered exponential describes
a non-uniform Gaussian random walk with mean $\mu_i$ and
covariance matrix $C_{i,j}$ :
\begin{eqnarray}
\mu_i &=&
- {\epsilon^3\over 24} \left( 2 - \sqrt{3} \right) v_i
\\
C_{i,j} &=& {\epsilon\over 2\sqrt{3}} \left[
\delta_{i,j} +
\left({1\over\sqrt{3}} - {1\over 2}\right)\epsilon^2 f_{i,j}
\right].
\end{eqnarray}
To sample this random distribution we need $\sqrt{C}$, which
we can approximate correctly to fourth order as
\begin{equation}
\left(\sqrt{C}\right)_{i,j} = 
 \sqrt{\epsilon\over 2\sqrt{3}}\left[
\delta_{i,j} + {1\over 2}
\left({1\over\sqrt{3}}-{1\over 2} \right)\epsilon^2 f_{i,j}
\right].
\end{equation}
Thus the entire factorization (\ref{dmc4p2}) can be
simulated by setting
\begin{eqnarray}
w_i &=& x_i +
     \xi_i\sqrt{{\epsilon\over 2} \left(1-{1\over\sqrt{3}}\right)},
     \nonumber \\
y_i &=& w_i(\epsilon/2)
     +\xi^\prime_i \sqrt{\epsilon\over 2 \sqrt{3}},\nonumber \\
z_i &=&
     y_i    - {\epsilon^3\over 24} \left( 2 - \sqrt{3} \right) v_i({\bf y})
	 + \sqrt{\epsilon\over 2 \sqrt{3}} 
   \left[
   \delta_{i,j} + {1\over 2}
   \left({1\over\sqrt{3}}-{1\over 2} \right)\epsilon^2 f_{i,j}({\bf y})
   \right]
   \xi_j^{\prime\prime},\nonumber\\
x_i^\prime &=& z_i(\epsilon/2) 
      +\xi^{\prime\prime\prime}_i
	  \sqrt{{\epsilon\over 2} \left(1-{1\over\sqrt{3}}\right)},
\label{dmc4al} 
\end{eqnarray}
where $\xi_i$ to $\xi_i^{\prime\prime\prime}$ are four sets of
independent Gaussian random numbers with zero mean and unit variance.  

As a severe test of the fourth order convergence of this algorithm, we use 
it to simulate the Brownian dynamics of 121 colloidal particles in two 
dimensions, with dimensionless surface density $N/A=0.5$, interacting 
via a pairwise strongly repulsive Yukawa potential
\begin{equation}
V(r)={V_0\over r}\exp[-\lambda(r-1)], 
\end{equation}
with $\lambda=8$. This system has
been described and simulated extensively via second order algorithms
by Branka and Heyes\cite{branka}. We will refer readers to this 
work for a detailed description of the system and their algorithms. 
In Fig. 1. we show the
convergence of the potential energy at one parameter setting as a function of 
the time step-size used. (Compare this figure to that of Fig. 6 of 
Branka and Heyes\cite{branka}.) The linear and quadratic convergences 
are clearly evident. The two second order algorithms used are as 
described by (\ref{altas}) and (\ref{altbs}). These are referred to in
Ref.\cite{chinatm} as algorithms LGV2b and LGV2a respectively. 

When our fourth order Langevin algorithm is implemented by using the standard 
fourth order Runge-Kutta algorithm to solve the trajectory equation (\ref{traject})
we obtained results as shown by open circles in Fig.1. The variance of the
potential energy increases abruptly at around $\epsilon=0.0028$ and
the algorithm becomes unstable at larger $\epsilon$'s. The problem can be
traced to the instability of the Runge-Kutta algorithm itself in solving for the
many-body dynamics.	While the trajectory evolution $\exp(\epsilon D)$ should
{\it always} decrease the potential energy,  
\begin{equation} 
{{dV}\over{dt}} ={{\partial V}\over{\partial{\bf x}}}
                 \cdot{{\partial{\bf x}}\over{\partial t}}= - |\nabla V|^2, 
\end{equation}
this is no longer respected by the Runge-Kutta algorithm at larger time steps.
The failure is due to the fact that
Gaussian random walks can deposit particles so
close together that the velocity field is changing too steeply for
the Runge-Kutta algorithm to integrate accurately.
Each of these particles then gets placed
chaotically somewhere in the periodic box, often again too near others,
thus multiplying the number of particles that will be moved
erratically in the next iteration. At time steps below but near
$\epsilon=0.0028$, the system can recover the regular behavior after
several to hundreds of iterations, but only at the cost of
increased variances and larger errors.
Thus the inaccuracy in the trajectory determination
causes the Langevin algorithm to fail prematurely.

To improve on this situation,
we monitor the difference between the results of the standard fourth order 
Runge-Kutta and the embedded second order algorithm (\ref{trk}).
We use the absolute value squared of this difference as a gauge of
the fourth order method, even though it is strictly only
an error estimate for the embedded second order algorithm.
If the value of this difference is larger than some tolerance
(0.01 in our case), we reject the result of the Runge-Kutta
and recompute the trajectory more accurately by applying our trajectory 
algorithm twice at half the time step size.
At small time steps, this incurs only a very small overhead. 
Even at a time step of $0.004$ only $3\%$ of the trajectories
have to be re-evaluated.
With this improvement, our fourth order Langevin algorithm gives results
as shown by solid circles in Fig. 1. (We also applied similar monitoring
processes to LGV2a and LGV2b by comparing the results of their first 
and second order Runge-Kutta algorithms.)
The step-size dependence of the fourth order algorithm is remarkably flat, 
and yielded the converged results of the lower algorithms at 
step-sizes nearly 50 times as large.

\section{Solving the Kramers Equation}

While we are not aware of other multivariable 4th order Langevin algorithms,
there are two fourth order algorithms in the literature for 
solving the Kramers equation in one dimension\cite{eli,drozdov}. Despite its 
more complicated appearance, the Kramers equation is actually simpler to solve 
than the Langevin equation. To illustrate the versatility of our 
operator approach, we will derive systematically a number of fourth order 
algorithms for solving this equation. Following Hershkovitz\cite{eli}, 
we write the Kramers equations in the form
\begin{equation}
\ddot{q}_i =F_i({\bf q}) - \gamma \dot{q}_i + \zeta_i,
\label{kram}
\end{equation}
where the force is derivable from a potential, 
$F_i({\bf q})=- \partial_i V({\bf q})$. A key simplification follows 
from the Hamilton form of the equation  
\begin{eqnarray}
\dot{q}_i &=& p_i \nonumber \\
\dot{p}_i &=& F_i({\bf q}) - \gamma {p}_i + \zeta_i,
\label{heq}
\end{eqnarray}
where $\zeta_i$ is the zero-mean Gaussian random noise vector with variance
\begin{equation}
\left< \zeta_i(t) \zeta_j(t^\prime) \right> = {2\over \beta}\gamma \delta_{ij} 
\delta(t-t^\prime).
\label{xinorm}
\end{equation}
The advantage here is that the noise only affects the momentum, and
classically, the momentum commutes with the position-dependent force term.
We will study the case of the bistable potential
\begin{equation}
V(q)=q^4-2q^2,
\label{pot}
\end{equation}
at parameter value $\gamma = 1$ and $\beta = 5$. For each algorithm considered
below, starting with $q(0)=0$ and $p(0)=0$, we evolve the system to 
a finite time of $t=6$. For comparison, we note that
the total energy approaches the equilibrium limit of $E=-0.8$ at infinite
time. 

Hershkovitz\cite{eli} has formally derived a 4th order algorithm for 
solving (\ref{heq}) using Taylor expansion, but he has given an explicit 
implementation only for one dimension.
In one dimension, each update of his algorithm requires one determination of the
particle trajectory to 4th order, 4 Gaussian random variables, and one evaluation
of the derivative of the force. 
The results of using his algorithm to evolve the system energy as a function of
the time step size $\epsilon$ is shown as solid squares in Fig.2. The 
standard 4th order Runge-Kutta algorithm, which requires four evaluations 
of the force, is used to solve for the particle's trajectory.    

To derive factorization algorithms in any dimension, we note that
the probability density function evolves according to
\begin{equation}
\dot P({\bf q},{\bf p},t) =L P({\bf q},{\bf p},t),
\end{equation}
where
\begin{equation}
L=
 {\gamma\over \beta} \nabla^2_{\bf p}
+ \gamma \nabla_{\bf p}\cdot{\bf p}
- {\bf p}\cdot \nabla_{\bf q} 
-{\bf F}({\bf q})\cdot\nabla_{\bf p} 
\equiv L_1+L_2+L_3+L_4.
\end{equation}
To factorize the evolution operator $\exp(\epsilon L)$ for small $\epsilon$, we
decompose $L$ into exactly solvable parts $T$ plus $D$ and apply known 
fourth order factorization schemes\cite{sufour,chinlang}. Drozdov and 
Brey\cite{drozdov} have recently initiated such a study of the 
Kramers equation. In this work, we have done 
an exhaustive search of all possible choices of solvable $T$ and $D$ 
such that $[D,[T,D]]$ or $[T,[D,T]$ is also solvable. We use the word 
``solvable" here loosely to denote either analytical result or trajectory
determination. For example, the effect of $\exp[\epsilon(L_2+L_3+L_4)]$ 
on the distribution function $P({\bf q},{\bf p},t)$ corresponds to
evolving the particle {\it trajectory} forward in time with a linear friction.
Since this can be computed using any trajectory integration algorithm, 
we consider $L_2+L_3+L_4$ to be solvable. While there are many solvable 
choices for $T$ and $D$, such as the sum of any two $L_i$, few resulting 
double commutators are simple. The possible 
choices for $T$ and $D$ are dramatically reduced if we insist that one of
their double commutators is also structurally similar to the original 
$T$ or $D$. 
There are then only three possibilities.

The first possibility is to take
\begin{eqnarray}
{T} &=& L_1+L_2+L_3,\nonumber\\
{D} &=& L_4,
\label{chh}
\end{eqnarray}
which is the choice originally made by Drozdov and Brey\cite{drozdov}. 
The Green's function corresponding to $\exp(\epsilon T)$ is known 
analytically\cite{drozdov}, and can be sampled via
\begin{eqnarray}
p_i^\prime &=& p_i{\rm e}^{-\gamma\epsilon}+\mu_i,
\nonumber\\
q_i^\prime &=& q_i+p_i(1-{\rm e}^{-\gamma\epsilon})/\gamma+\nu_i,
\label{pqup}
\end{eqnarray}
where corresponding to each pair of $(p_i,q_i)$, $(\mu_i,\nu_i)$ is a pair of
correlated Gaussian random numbers given by
\begin{eqnarray}
\mu_i&=& \xi_i\sqrt{{1\over\beta}\Bigl( 1-{\rm e}^{-2\gamma\epsilon}\Bigr)},
\nonumber\\
\nu_i&=&{1\over{\gamma}}
\left ( 
{1-{\rm e}^{-\gamma\epsilon}}\over{1+{\rm e}^{-\gamma\epsilon}}
\right )\mu_i
+\xi_i^\prime
\sqrt{ {1\over{\beta\gamma^2}}\left( 2\gamma\epsilon-4\left ( 
{1-{\rm e}^{-\gamma\epsilon}}\over{1+{\rm e}^{-\gamma\epsilon}}
\right )\right)}.
\label{mnup}
\end{eqnarray}
Here, $\xi_i$ and $\xi_i^\prime$ are again two independent Gaussian 
random numbers with zero mean and unit variance. Note that at a given
step size $\epsilon$, all the above functions involving 
${\rm e}^{-\gamma\epsilon}$, etc., only need to be evaluated once at
the beginning of the simulation.  
The operator $\exp(\epsilon D)$ can be exactly simulated by
\begin{equation}
p\,^\prime_i=p\,_i+\epsilon F_i({\bf q}).
\label{momen}
\end{equation}
As we will see, this choice is clever because there is no trajectory 
equation to solve. The double commutator required for a fourth order 
factorization is
\begin{equation}
[D,[T,D]]=[L_4,[L_3,L_4]]=-\nabla_{\bf q}|{\bf F}|^2\cdot\nabla_{\bf p} 
\label{dfour}
\end{equation}
which is just $D$ but with a force $\nabla_{\bf q}|{\bf F}|^2$.
For each choice of $T$ and $D$, there are three generic 
schemes\cite{chinlang} for factorizing the decomposed operator 
$\exp[\epsilon(T+D)]$ to fourth order with purely positive
coefficients. For this choice of $T$ and $D$,
we found that schemes A and B of Ref.\cite{chinlang} give rather similar 
results, so we will only
present results for schemes A and C. Scheme A and C are respectively,  
\begin{equation}
{\rm e}^{\epsilon (T + D) }
=
{\rm e}^{{1\over 6}\epsilon  {D}}
{\rm e}^{{1\over 2}\epsilon  {T}}
{\rm e}^{{2\over 3}\epsilon  \widetilde D}
{\rm e}^{{1\over 2}\epsilon  {T}}
{\rm e}^{{1\over 6}\epsilon  {D}}
+O(\epsilon ^5),
\label{kfoura}
\end{equation}
and 
\begin{equation}
{\rm e}^{\epsilon (T + D) }
=
{\rm e}^{{1\over 6}\epsilon  {T}}
{\rm e}^{{3\over 8}\epsilon  {D}}
{\rm e}^{{1\over 3}\epsilon  {T}}
{\rm e}^{{1\over 4}\epsilon  \widetilde D}
{\rm e}^{{1\over 3}\epsilon  {T}}
{\rm e}^{{3\over 8}\epsilon  {D}}
{\rm e}^{{1\over 6}\epsilon  {T}}
+O(\epsilon ^5),
\label{kfourc}
\end{equation}
where
\begin{equation}
\widetilde D
=D+{\epsilon^2\over{48}}[D,[T,D]].
\label{tilded}
\end{equation}
The results of these two algorithms are shown as solid and open
circles in Fig.2. We will refer to these two as algorithms DB (Drozdov and Brey)
and K4a respectively. Each algorithm evaluates the force three times 
and the derivative of the force once. Drozdov and Brey's algorithm uses 4 Gaussian
random numbers and K4a uses eight. For the extra effort, algorithm K4a has a 
much flatter convergence curve. Drozdov and Brey solved their one dimensional 
problem on a grid. We used Monte Carlo simulation, which can be generalized to
any dimension.   

The second possibility is to take
\begin{eqnarray}
{T} &=& L_1+L_2,\\
{D} &=& L_3+L_4.
\label{chtwo}
\end{eqnarray}
The operator $\exp(\epsilon T)$ now corresponds to an Ornstein-Uhlenbeck process 
in $p_i$\,,
\begin{equation}
p_i^\prime = p_i{\rm e}^{-\gamma\epsilon}
+\xi_i\sqrt{{1\over\beta}\Bigl( 1-{\rm e}^{-2\gamma\epsilon}\Bigr)},
\label{pou}
\end{equation}
and $\exp(\epsilon D)$ evolves the particle trajectory 
forward in time without friction,
\begin{eqnarray}
p_i^\prime &=& p_i(\epsilon),
\nonumber\\
q_i^\prime &=& q_i(\epsilon).
\label{pqsec}
\end{eqnarray}
In this case, the simpler double commutator is
\begin{equation}
[T,[D,T]]=[L_2,[D,L_2]]=-\gamma^2 D,
\label{bfour}
\end{equation}
which {\it does not} require the derivative of the force.
For this choice, we need to switch $T\leftrightarrow D$ in
scheme A and slightly modify it as follows: 
\begin{equation}
{\rm e}^{\epsilon (T + D) }
=
{\rm e}^{ {1\over 6}\epsilon T}
{\rm e}^{ {1\over 2}\epsilon [1 - \epsilon^2\gamma^2/72] D}
{\rm e}^{ {2\over 3}\epsilon T}
{\rm e}^{ {1\over 2}\epsilon [1 - \epsilon^2\gamma^2/72] D}
{\rm e}^{ {1\over 6}\epsilon T}
+O(\epsilon^5).
\end{equation}
The effect of the double commutator simply reduces the time of the
trajectory evolution. This algorithm, which will be referred to as K4b, 
requires two trajectory determinations but no derivative of the force and
only three Gaussian random numbers. The trajectory can be computed using
the standard 4th order Runge-Kutta algorithm with four force evaluations, or the
4th order Forest-Ruth symplectic algorithm\cite{forest} with three force evaluations.
The results from these two cases are plotted as solid and open diamonds 
respectively in Fig.2.
For this choice of $D$, we did not bother with factorization schemes B 
or C, since either would have required more than two trajectory determinations.    

The third possibility is to take
\begin{eqnarray}
{T} &=& L_1,\\
{D} &=& L_2+L_3+L_4,
\label{chthree}
\end{eqnarray}
where now $\exp(\epsilon T)$ is just a Gaussian process 
in $p_i$,
\begin{equation}
p_i^\prime = p_i +\zeta_i\sqrt{\epsilon},
\label{pthird}
\end{equation}
and $\exp(\epsilon D)$ evolves the particle trajectory 
forward in time with friction. For this case, we have the simplest result,
\begin{equation}
[T,[D,T]]=0,
\label{cfour}
\end{equation}
and a simplified fourth order factorization 
\begin{equation}
{\rm e}^{\epsilon (T + D) }
=
{\rm e}^{ {1\over 6}\epsilon T}
{\rm e}^{ {1\over 2}\epsilon D}
{\rm e}^{ {2\over 3}\epsilon T}
{\rm e}^{ {1\over 2}\epsilon D}
{\rm e}^{ {1\over 6}\epsilon T}
+O(\epsilon^5).
\end{equation}
We shall refer to this as algorithm K4c. This algorithm is similar to K4b, with
no force derivative necessary. If we solve the trajectory equation by
the 4th order Runge-Kutta algorithm, we obtain results as shown by 
solid triangles in Fig.2. Note that in contrast to previous algorithms, this
algorithm does not converge monotonically. It overshoots and converges from the
top.

In the course of our calculations, we find that for each algorithm, a more 
accurately determined particle trajectory will yield a flatter convergence
curve. If we now further decompose $D=D_1+D_2$ in algorithm K4c, with
\begin{eqnarray}
D_1 &=& L_2,\nonumber\\
D_2 &=& L_3+L_4,
\end{eqnarray}
the double commutator $[D_1,[D_2,D_1]]=-\gamma^2 D_2$ is just
a restatement of (\ref{bfour}). We can again factorize,
\begin{equation}
{\rm e}^{\epsilon D}
=
{\rm e}^{ {1\over 6}\epsilon D_1}
{\rm e}^{ {1\over 2}\epsilon [1 - \epsilon^2\gamma^2/72] D_2}
{\rm e}^{ {2\over 3}\epsilon D_1}
{\rm e}^{ {1\over 2}\epsilon [1 - \epsilon^2\gamma^2/72] D_2}
{\rm e}^{ {1\over 6}\epsilon D_1}
+O(\epsilon^5).
\end{equation}
The friction evolution ${\rm e}^{\epsilon D_1}$ rescales the momentum, 
\begin{equation}
p_i^\prime = p_i {\rm e}^{-\gamma\epsilon},
\end{equation}
and ${\rm e}^{\epsilon D_2}$ again evolves the trajectory forward for time 
$\epsilon$.
This way of solving the trajectory with friction doubles the number of trajectory
calculations, but also further flattens the convergence curve. To minimize the
number of force evaluations, we use the Forest-Ruth symplectic algorithm to calculate
the trajectory. The results are shown as open triangles in Fig.2.

Of the algorithms studied, Drozdov and Brey's algorithm makes maximum use of analytical
knowledge and is very efficient. The improvement we suggested, algorithm K4a, with
twice the number of Gaussian random numbers, seemed to double the range of the 
convergence. Our new algorithms K4b and K4c, while requiring two trajectory
determinations, have no need of evaluating the force derivative. All these algorithms
serve to illustrate the power of the factorization method. While the diligence of
Hershkovitz is rewarded with just a single fourth order algorithm, 
we can survey the form of the evolution operator and derive
many fourth order algorithms.

\section{Summary and Conclusions}

In this work, we have shown how the method of operator factorization can be 
applied to the Langevin equation to derive a practical fourth order 
algorithm. This method of factorizing an evolution 
operator of the form ${\rm e}^{\epsilon(A+B)}$ leads to {\it unitary} algorithms 
for solving the Schr$\ddot{\rm o}$dinger equation in quantum mechanics, 
{\it symplectic} algorithms for solving Hamilton's equations in classical 
mechanics, and {\it norm preserving} algorithms for solving the Langevin 
equation in stochastical mechanics. A key step in deriving a fourth order
Langevin algorithm is our treatment of the double commutator term through
successive use of normal ordering. The resulting algorithm (\ref{dmc4al}) is
computationally demanding, but one is rewarded by a very flat convergence 
curve, virtually eliminating the step-size dependent error. Future use of this
algorithm in other applications may lead to further simplifications and 
enhancements of its utility.

We also derived a number of 4th order algorithms for solving the Kramers 
equation. The freedom in decomposing the kernel operator and choosing a 
particular factorization scheme illustrates the power of this approach. 
It is difficult to see these global structures from just doing Taylor expansions.
One advantage of our simulation approach is that we are not restricted to 
solving the Kramers equation in one dimension. We can solve it in any dimension. 
Our use of the Kramers equation is also only illustrative, one can apply this
method of operator factorization to other stochastic equations of one's
own interest.

It is observed in solving both equations that the step-size error is reduced by
solving the trajectory more exactly. Different fourth order algorithms for 
solving the trajectory equation can yield different convergence curves.
One should therefore explore the effect of using fourth order algorithms
other than Runge-Kutta in implementing any of the above stochastic algorithms.

\acknowledgements
This research was funded, in part, by the U. S. National Science Foundation 
grants PHY-9512428, PHY-9870054 and DMR-9509743.


\ifpreprintsty\newpage\fi
\ifpreprintsty\newpage\fi
\begin{figure}
\noindent
\vglue 0.2truein
\hbox{
\vbox{\hsize=7truein
\epsfxsize=6truein
\leftline{\epsffile{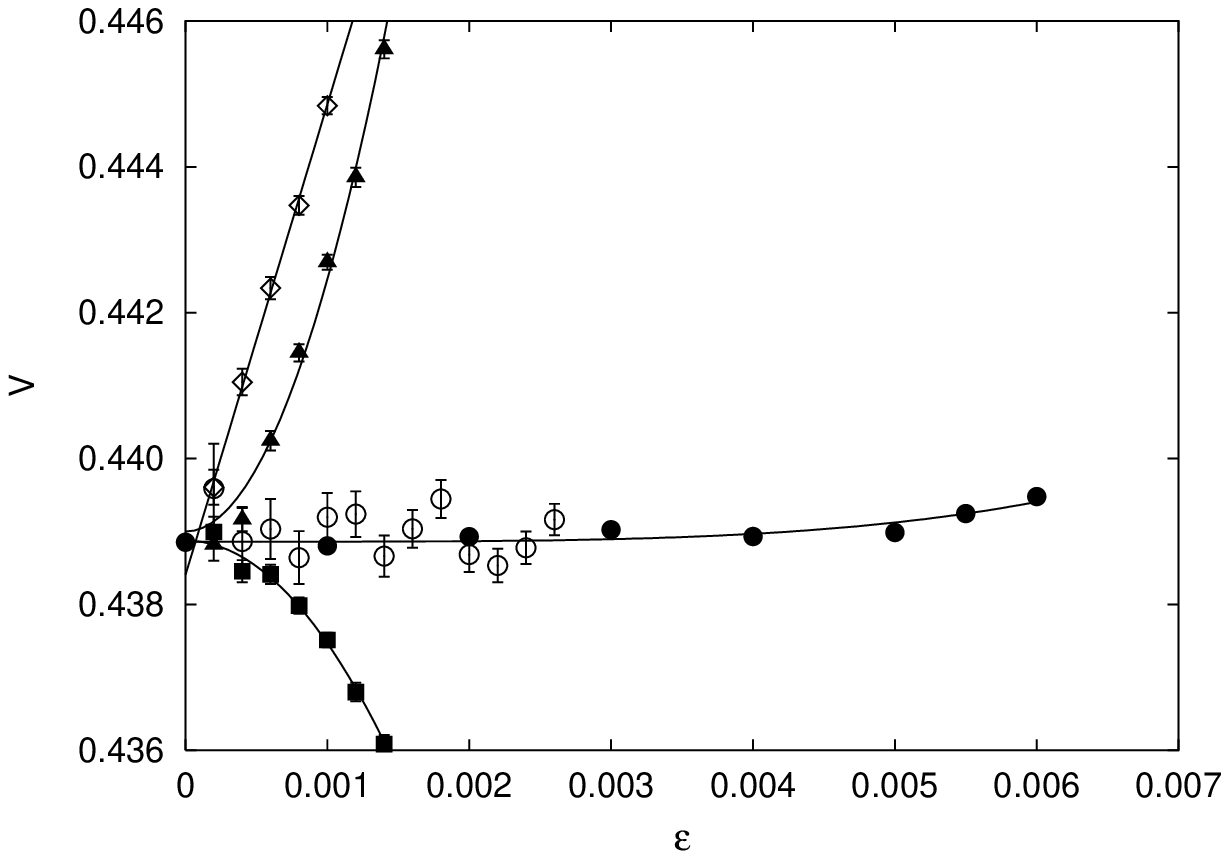}}
}
}
\vglue 0.3truein
\caption{The convergence of
Langevin algorithms for simulating the Brownian dynamics of
121 interacting colloidal particles in two dimensions. The equilibrium
potential energy per particle is plotted as a function of the time step size
$\epsilon$ used. Open diamonds are results using the first order Langevin
algorithm. Solid triangles and solid squares denote results of the two second
order algorithms LGV2a and LGV2b respectively, as described in the text.
Open circles give results of our fourth order Langevin algorithm using the
standard 4th order Runge-Kutta algorithm for determining the particle
trajectory. The solid circles give results with improved trajectory
determination as discussed in the text. 
}
\label{fone}
\end{figure}
\ifpreprintsty\newpage\fi
\begin{figure}
\noindent
\vglue 0.2truein
\hbox{
\vbox{\hsize=7truein
\epsfxsize=6truein
\leftline{\epsffile{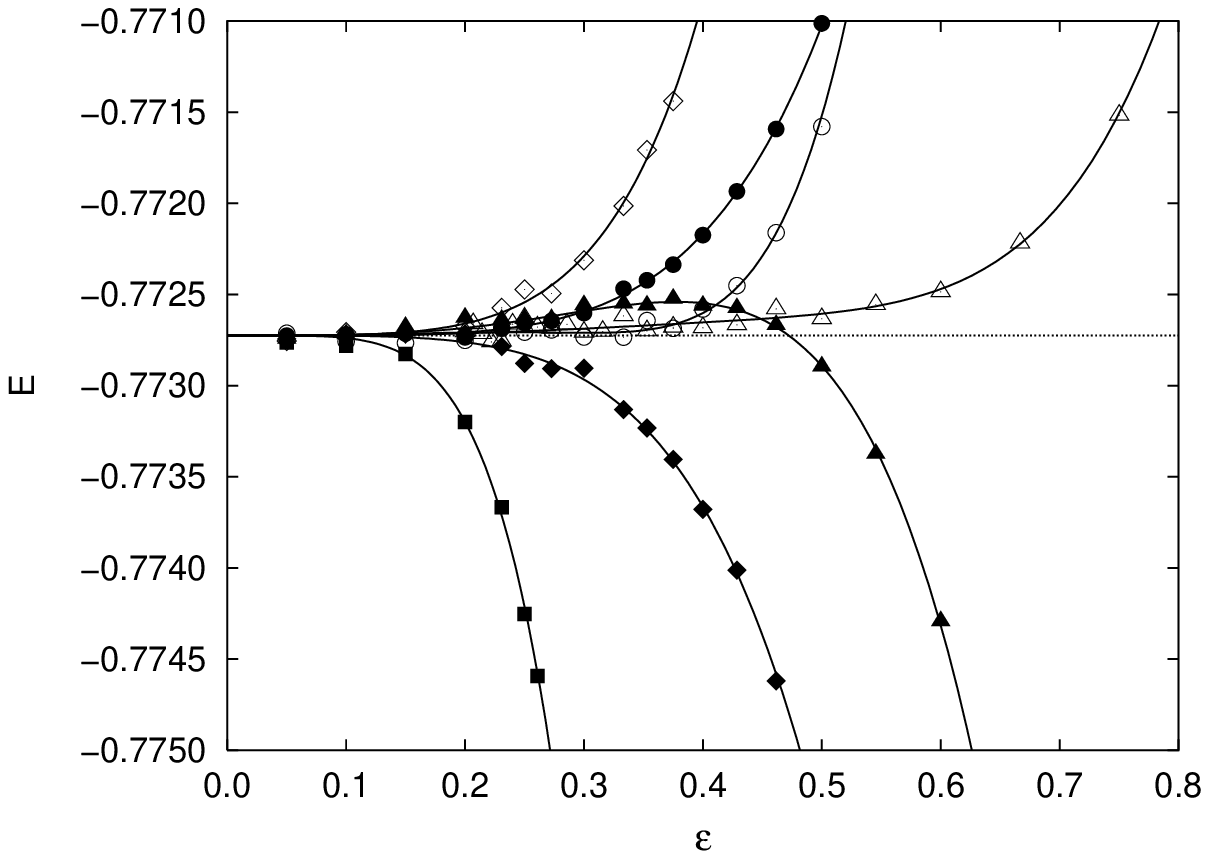}}
}
}
\vglue 0.3truein
\caption{The convergence of various fourth order algorithms for
solving the Kramers equation in one dimension. The energy calculated
is at a finite time of $t=6$ with system parameters
$\beta=5$ and $\gamma=1$. 
Solid squares: Hershkovitz's algorithm. Solid and open circles:
Drozdov and Brey's algorithm and K4a. Solid and open diamonds: two variants of
algorithm K4b. Solid and open triangles: two variants of algorithm K4c. 
See text for algorithm descriptions. The fitted lines 
all have leading term $\epsilon^4$ or higher. 
Error bars are comparable or smaller than the size of
plotting symbols.
}
\label{ftwo}
\end{figure}

\begin{thebibliography}{10}

\bibitem{risken}
H. Risken, {\it The Fokker-Planck Equation, Methods of Solution
and Applications}, 2nd edition, (Springer, New York, 1989).

\bibitem{helf}
E. Helfand. Bell Syst. Tech. J., {\bf 58}, 2289 (1979).

\bibitem{drum}
I. Drummond {\it et. al.}, Nucl. Phys. {\bf B220}, 119 (1983).

\bibitem{ukawa}
A. Ukawa and M. Fukugita, Phys. Rev. Lett. {\bf 55}, 1854 (1985).
 
\bibitem{chinnp}
S. A. Chin, Nucl. Phys. {\bf B (Proc. Suppl.) 9}, 498 (1989).

\bibitem{chinatm}
S. A. Chin, Phys. Rev. {\bf A42}, 6991 (1990).
  
\bibitem{eli}
E. Hershkovitz, J. Chem. Phys. {\bf 108}, 9253 (1998).

\bibitem{sufour}
M. Suzuki, {\it Computer Simulation Studies in Condensed 
Matter Physics VIII},
eds, D. Landau, K. Mon and H. Shuttler (Springler, Berlin, 1996).

\bibitem{chinlang}
S. A. Chin, Phys. Lett. {\bf A226}, 344 (1997) .

\bibitem{branka}
A. Branka and D. Heyes Phys. Rev. {\bf E60}, 2381 (1999) . 
 
\bibitem{drozdov}
A. N. Drozdov and J. J. Brey, Phys. Rev. {\bf E57}, 1284 (1998) .


\bibitem{nogo}
M. Suzuki, J. Math. Phys. {\bf 32}, 400 (1991).


\bibitem{forest}
E. Forest and R. D. Ruth, Physica D {\bf 43}(1990) 105.







\end{thebibliography}
\end{document}